\documentclass[pra,10pt,twocolumn,floatfix,amsmath,showpacs,superscriptaddress]{revtex4-1}
\usepackage[ansinew]{inputenc}

\usepackage{verbatim}
\usepackage{epsfig}
\usepackage{pst-all}
\usepackage{color}
\usepackage{dcolumn}
\usepackage{amsmath}
\usepackage{amsthm}
\usepackage{bm}
\usepackage{layout}
\usepackage{float}
\usepackage{txfonts}
\usepackage{amsfonts}
\usepackage{amssymb}%
\usepackage[ansinew]{inputenc}
\usepackage{amsmath}
\usepackage{amsthm}
\usepackage{float}
\usepackage{amsfonts}
\usepackage{amssymb}

\def\br{\mathbf{r}}

\def\ket#1{|#1\rangle}
\def\bra#1{\langle#1|}

\newtheorem{lemma}{Lemma}

\newtheorem{definition}{Definition}
\newtheorem{proposition}[definition]{Proposition}

\newtheorem{theorem}[definition]{Theorem}
\newtheorem{corollary}[definition]{Corollary}
\newtheorem{conjecture}[definition]{Conjecture}

\newtheorem{remark}[definition]{Remark}
\newtheorem{example}[definition]{Example}
\newtheorem{question}[definition]{Question}

\def\ra{\rightarrow}
\def\a{\alpha}
\newcommand{\ketbra}[2]{|#1\rangle\!\langle#2|}

\def\sm{\setminus}
\def\dg{\dagger}
\def\p{\pi}
\def\r{\rho}

\newcommand{\cH}{{\cal H}}
\def\ox{\otimes}
\def\tr{\mathop{\rm Tr}}
\def\bcj{\begin{conjecture}}
\def\ecj{\end{conjecture}}
\def\bcr{\begin{corollary}}
\def\ecr{\end{corollary}}
\def\bd{\begin{definition}}
\def\ed{\end{definition}}
\def\bea{\begin{eqnarray}}
\def\eea{\end{eqnarray}}
\def\bem{\begin{enumerate}}
\def\eem{\end{enumerate}}
\def\bex{\begin{example}}
\def\eex{\end{example}}
\def\bim{\begin{itemize}}
\def\eim{\end{itemize}}
\def\bl{\begin{lemma}}
\def\el{\end{lemma}}
\def\bpf{\begin{proof}}
\def\epf{\end{proof}}
\def\bpp{\begin{proposition}}
\def\epp{\end{proposition}}
\def\bqu{\begin{question}}
\def\equ{\end{question}}
\def\br{\begin{remark}}
\def\er{\end{remark}}
\def\bt{\begin{theorem}}
\def\et{\end{theorem}}
\newcommand{\proj}[1]{| #1\rangle\!\langle #1 |}
\def\ph{\varphi}
\def\min{\mathop{\rm min}}

\newcommand{\jpa}{J. Phys. A}

\begin{document}

\title{Non-zero total correlation means non-zero quantum correlation}

\author{Bo Li}
\affiliation{Department of Mathematics and
Computer, Shangrao Normal University, Shangrao 334001, China\\ and Institute of Physics, Chinese Academy of Sciences,
Beijing 100190, China}

\author{Lin Chen}
\email{linchen0529@gmail.com} \affiliation{Department of Pure
Mathematics, University of Waterloo,
  Waterloo, Ontario, Canada\\Institute for Quantum Computing, University of
Waterloo,
  Waterloo, Ontario, Canada\\
  and Center for Quantum Technologies, National University of
Singapore, Singapore}

\author{Heng Fan}
\affiliation{Institute of Physics, Chinese Academy of Sciences,
Beijing 100190, China}

\date{\today}

\begin{abstract}
We investigate the super quantum discord based on weak measurement.
It is an extension of the standard quantum discord defined by
projective measurement and still describes the quantumness of
correlations. We provide a few equivalent conditions on zero super
quantum discord by using quantum discord, classical correlation and
mutual information. In particular, we find that super quantum
discord is zero only for product states, which meantime has zero
mutual information. This provides a perspective that non-zero
correlations can always be viewed by quantum correlation with weak
measurement. As an application, we present the assisted state
discrimination scheme.
\end{abstract}


\maketitle


\section{Introduction}\label{intr}

Quantum measurement plays a key role in quantum mechanics. It has
some interesting quantum properties that are rarely seen in daily
life. The properties include the collapse of wavefunction,
compatible observables and the contextuality phenomena. To realize a
quantum measurement, one need to construct a set of orthogonal
projection operators  corresponding to the observable eigenvector
spaces of a Hermitian operator. The possible outcomes of the
measurement correspond to the eigenvalues of the Hermitian operator.
This is the standard von Neumann measurement or projective
measurement \cite{neumann}. Recently, the  formalism was generalized
to the positive operator valued measure (POVM)
\cite{Nielsen&Chuang}. It can really capture many phenomena beyond
projective measurement.

However, the measurement of quantum state inevitably disturbs the
quantum system which in turn determines our retrieved knowledge
about the measured system. In order to make the least influence on
original quantum state, one may introduce a measurement that induces
a partial collapse of a quantum state. This is the so-called weak
measurement \cite{aav,Korotkiv2006,Sun2009}. Quantum states can be
retrieved with a nonzero success probability when the interaction
between the system and measurement apparatus is weak
\cite{Ueda1992}. It has been shown that any generalized measurement
can be decomposed into a sequence of weak measurements \cite{brun}.
Therefore weak measurement is universal. The reverse process has
also attracted much attention both theoretically and experimentally
\cite{wise1,mir}, due to its potential applications in quantum
information processing \cite{Korotkov2010}. Weak measurement can
also amplify very tiny signals \cite{kwiat,lundeen}.

Searching for quantum correlation in composite system and
identifying its role in quantum information processing is one of the
fundamental problems in quantum mechanics. Quantum entanglement is
extensively regarded as a crucial role in quantum teleportation and
super dense coding, etc \cite{Nielsen&Chuang}. Quantum discord
\cite{hv01,oz02,modi}, which is beyond quantum entanglement, can
effectively grasp the role of quantumness of correlations and is
different from the classical correlation. It is shown to be present
in deterministic quantum computation with one qubit (DQC1)
\cite{Knill}, as a resource in remote state preparation \cite{dak}.
Also, the consumed discord bounds the quantum advantage in encode
information \cite{gu}. Quantum dissonance (or one side discord) is
shown to be required in optimal assisted discrimination
\cite{Roa,bo1,fulin}. We know that quantum entanglement can be
described and detected by various methods \cite{peres1,hhh96,cd12}.
Quantum discord, on the other hand, can exist when entanglement is
absent. It vanishes for the so-called classical-classical (CC)
state, the classical-quantum (CQ) state or the quantum-classical
(QC) state \cite{datta08,dvb10,ccm11}.

However, evidences show that quantum advantage may still exist even
for vanishing discord \cite{Bennett2011}. It is then expected to
construct a measure of quantum correlation which always exists
except for product states. A good candidate for this measure is the
super quantum discord. This is an extension of quantum discord with
weak measurement \cite{sp12}. It is found that super quantum discord
can capture much more quantum correlation in the sense that the
super quantum discord is always larger than the normal discord
induced by the strong (projective) measurement. Furthermore, super
discord can result in an improvement on the entropic uncertainty
relations \cite{dd,ren}. Now we ask, what is the criterion by which
super quantum discord exists in quantum system? Can super discord
exist in some quantum information model where quantum discord and
entanglement do not exist? In this paper, we provide a necessary and
sufficient condition for vanishing super discord in terms of
classical correlation, mutual information, and normal discord. Our
results show that quantum correlation measured by super quantum
discord always exist except there is no correlation. Thus we can
confirm the expectation that all correlations can be viewed from
perspective of quantum correlation. We further illustrate that super
discord can present in optimal assisted state discrimination on both
sides, where only one side of quantum discord is present and
entanglement is totally not needed.

This paper is organized as follows. In Sec. \ref{prelimin}, we
recall some definition and property of super discord. In Sec.
\ref{consuper}, we provide a series of necessary and sufficient
condition on vanishing super discord. An illustration of super
discord present on both sides in optimal assisted state
discrimination is given in In Sec. \ref{sudiscrimina}. Finally we
summary in Sec. \ref{summ}.

\section{the concept and property of super discord}\label{prelimin}

Consider the bipartite state $\r$ on the space $\cH_A\ox\cH_B$. Let
$\{\pi_k\}$ be one-dimensional von Neumann projectors, and the
probability $p_k=\tr (I\ox\pi_k) \r (I\ox\pi_k)$. The completeness
of the operators $\{\pi_k\}$ implies the formula $\sum_k p_k=1$.
Next, $S(\r):=-\tr \r \log \r$ is the von Neumann entropy and
``$\log$" denotes ``$\log_2$" throughout the paper. We refer to
$\r_A,\r_B$ as the reduced density operators of $\r$. Then we denote
$I(\r):=S(\r_A)+S(\r_B)-S(\r)$ as the mutual quantum information and
$C(\r):=\max_{\pi_k} I \bigg(\sum_k (I\ox\pi_k) \r (I\ox\pi_k)
\bigg)$ as the classical correlation \cite{hv01,oz02,ll11}. Both of
them are non-negative because the mutual information is non-negative
\cite{Nielsen&Chuang}.

The quantum discord for $\r$ is defined as the difference between
the mutual information and classical correlation \cite{oz02,datta08}
 \bea
 \label{ea:discord}
  D(\r)
  &=&I(\r)-C(\r)
  \notag\\
  &=&
  S(\r_B) - S(\r) + {\min}_{\pi_k}
  \sum_k p_k S \bigg( \frac{(I\ox\pi_k) \r (I\ox\pi_k)}{p_k} \bigg)
  .
 \eea
It is known that \cite{dvb10,ccm11} the (``right'') discord is zero if and only
if $\r=\sum_i p_i \r_i\ox\proj{\ph_i}$, where the $\ket{\ph_i}$ are
o. n. basis. This is the so-called classical state in the system
$B$.

Next we recall the super quantum discord $D_w(\r)$ for two-qubit
states $\r$ introduced in \cite{sp12}. It is defined as
 \bea
 \label{ea:superDISCORD}
D_w(\r) := \min_{ \{\p_0,\p_1\} } S_w (A|\{ P^B(x) \}) - S(A|B),
 \eea
where the conditional entropy $S(A|B) = S(\r) - S(\r_B)$ and
 \bea
 \label{ea:Sw}
 S_w (A|\{ P^B(x) \}) &=& p(x) S (\r_{ A | P^B(x) }) + p(-x) S (\r_{ A | P^B(-x) }),
 \\
 \label{ea:p(pm x)}
 p(\pm x) &=& \tr \bigg( (I\ox P^B(\pm x)) \r (I\ox P^B(\pm x)) \bigg),
 \\
 \label{ea:APB(pm x)}
 \r_{ A | P^B(\pm x) } &=& {1 \over p(\pm x)} \tr_B \bigg ( (I\ox P^B(\pm x)) \r (I\ox P^B(\pm x)) \bigg),
 \\
 \label{ea:P(x)}
 P(x) &=&
 \sqrt{1-\tanh x \over 2} \p_0
 +
 \sqrt{1+\tanh x \over 2} \p_1,
 \\
 \label{ea:P(-x)}
 P(-x) &=&
 \sqrt{1+\tanh x \over 2} \p_0
 +
 \sqrt{1-\tanh x \over 2} \p_1,
 \eea
and $x\in R\sm\{0\}$ is a parameter describing the strength of
measurement process. By Eq. \eqref{ea:superDISCORD}, we have
$D_w(U\ox V\r U^\dg \ox V^\dg)\le D_w(\r)$ with arbitrary unitary
$U,V$. One may similarly obtain $D_w(U\ox V\r U^\dg \ox V^\dg)\ge
D_w(\r)$, so we have
 \bea
 \label{ea:LUinvariant}
D_w(U\ox V\r U^\dg \ox V^\dg)=D_w(\r).
 \eea
That is, the super discord is invariant up to the local unitary.
This property is the same as that of normal discord.

By Eqs. \eqref{ea:P(x)} and \eqref{ea:P(-x)}, we obtain the
completeness relation
 \bea
 \label{ea:p0+p1}
 \p_0+\p_1 &=& P(x)^\dg P(x) + P(-x)^\dg P(-x) = I.
 \eea
By Eqs. \eqref{ea:p(pm x)} and \eqref{ea:p0+p1}, we see that the
probability sum is equal to one:
 \bea
 \label{ea:p(x)+p(-x)}
p(x)+p(-x)=1.
 \eea
Using the concavity of von Neumann entropy and Eqs. \eqref{ea:Sw}
and \eqref{ea:APB(pm x)}, we easily obtain $I(\r)\ge D_w(\r)$. By
combining the Theorem of \cite{sp12}, we have
 \bea
 \label{ea:super>=normal}
 I(\r) \ge D_w(\r) \ge D(\r)
 \eea
for any two-qubit states. However these three quantities are not
quantitatively related to the classical correlation. Indeed, it
follows from Ref. \cite{luo08} that the difference $C(\r)-D(\r)$
can be either positive or negative for two-qubit Bell diagonal
states $\r$ in \cite{luo08}, see also \cite{lwf11}.
Nevertheless, we will determine the relations between classical
correlation, mutual information, super discord and discord for the
product states in next section.

\section{condition for zero super discord}\label{consuper}

Similar to the case of discord we ask the following question: what
are the states $\r$ whose super discord is zero? By Eq.
\eqref{ea:super>=normal} and \cite{dvb10,ccm11}, such states $\r$
must be classical in the system $B$. However the converse is not
evidently true, see Theorem \ref{thm:zeroSUPERdiscord}. For this
purpose we need a preliminary lemma. It is known that the classical
correlation is zero for the product state \cite{hv01}. We show that
the inverse is also true.
 \bl
 \label{le:C(rho)=0}
Any bipartite state $\r$ realizing $C(\r)=0$ is a product state,
i.e., $\r=\r_A\ox\r_B$.
 \el
 \bpf
By definition, the condition $C(\r)=0$ implies that $I \bigg(\sum_k
(I\ox\pi_k) \r (I\ox\pi_k) \bigg)=0$ holds for any $\{\p_k\}$. By
the subadditivity of von Neumann entropy, the state $\sum_k
(I\ox\pi_k) \r (I\ox\pi_k)$ is a product state. By tracing out the
system $A$ or $B$, we have
 \bea
 \label{ea:C(rho)}
 \sum_k (I\ox\pi_k) \r (I\ox\pi_k)
 =
 \r_A \ox \sum_k \p_k \r_B \p_k
 \eea
for any $\{\p_k\}$. Let $\r_B=\sum_i p_i \proj{b_i}$ be the spectral
decomposition, and we can assume $\r=\sum_{ij} \r_{ij} \ox
\ketbra{b_i}{b_j}$. By choosing $\p_i=\proj{b_i}$ in Eq.
\eqref{ea:C(rho)}, we obtain $\r_{ii}=p_i \r_A$, $\forall i$. Using
the normalization condition $\sum_i p_i=1$ we have
$\r=\r_A\ox\r_B+\sum_{i\ne j} \r_{ij} \ox \ketbra{b_i}{b_j}$. By
using $\r$ in Eq. \eqref{ea:C(rho)}, we have
 \bea
 \label{ea:C(rho)2}
 \sum_k (I\ox\pi_k) \bigg( \sum_{i\ne j} \r_{ij} \ox \ketbra{b_i}{b_j} \bigg) (I\ox\pi_k)
 =
 0
 \eea
for any $\{\p_k\}$. Since any two summands are orthogonal, we have
$(I\ox\pi_k) \bigg( \sum_{i\ne j} \r_{ij} \ox \ketbra{b_i}{b_j}
\bigg) (I\ox\pi_k)=0$, $\forall k$. By choosing
 \bea
 \p_k &=&
 \bigg( \frac{1}{\sqrt2}\ket{b_l}+\frac{1}{\sqrt2}e^{i\a}\ket{b_j} \bigg)
 \bigg( \frac{1}{\sqrt2}\bra{b_l}+\frac{1}{\sqrt2}e^{-i\a}\bra{b_j} \bigg),
 \eea
we have $\r_{lj}e^{i\a}+\r_{jl}e^{-i\a}=0$ for any real $\a$. So
$\r_{lj}=\r_{jl}=0$ for any $j\ne l$. Thus $\r=\r_A\ox\r_B$ and the
assertion follows. This completes the proof.
 \epf

\begin{theorem}
 \label{thm:zeroSUPERdiscord}
The following seven statements are equivalent for the two-qubit
state $\r$:
\\ (a) $\r$ is a product state;
\\ (b) $\r$ has zero classical
correlation;
\\ (c) $\r$ has zero super discord;
\\ (d) $\r$ has zero mutual information;
\\ (e) $\r$ has equal discord and super discord;
\\ (f) $\r$ has equal discord and mutual information;
\\ (g) $\r$ has equal super discord and mutual information.
\end{theorem}
 \bpf
\textit{(a)$\ra$(b)} follows from the definition of classical
correlation.

\textit{(b)$\ra$(c)} By Lemma \ref{le:C(rho)=0} we may suppose
$\r=\r_A\ox\r_B$. By Eqs. \eqref{ea:p(pm x)} and \eqref{ea:APB(pm
x)}, we obtain $ \r_{ A | P^B(\pm x) }=\r_A$. By Eqs. \eqref{ea:Sw}
and \eqref{ea:p(x)+p(-x)}, we have $S_w (A|\{ P^B(x) \}) = S(\r_A)$.
Then Eq. \eqref{ea:superDISCORD} implies that $D_w(\r)=0$, so
(b)$\ra$(c) follows.

\textit{(c)$\ra$(e)} follows from $D(\r)\ge0$ and Eq.
\eqref{ea:super>=normal}.

\textit{(e)$\ra$(d)}. Let $\{\p_i\}$ be the measurement basis that
minimizes the super discord in Eq. \eqref{ea:superDISCORD}. By
\cite[Eq. (11)]{sp12} and Eq. \eqref{ea:discord}, we have
 \bea
 \label{ea:inequality}
 D_w(\r)
 &\ge &\sum^1_{k=0} p_k S \bigg( \frac{(I\ox\p_k) \r (I\ox\p_k)}{p_k} \bigg)- S(A|B)
 \nonumber \\
 &\ge &D(\r).
 \eea
By the hypothesis, both equalities in Eq. \eqref{ea:inequality}
holds. By results in \cite{sp12} and the concavity of von Neumann
entropy, the first equality holds only if $\tr_B\frac{(I\ox\p_0) \r
(I\ox\p_0)}{p_0}=\tr_B \frac{(I\ox\p_1) \r (I\ox\p_1)}{p_1}$
\cite{Nielsen&Chuang}. Since the operators $\pi_k$ are of rank one, the second
equality implies
 \bea
 \label{ea:necessity}
 D(\r)
 &=& \sum^1_{k=0} p_k S \bigg( \tr_B \frac{(I\ox\p_k) \r
(I\ox\p_k)}{p_k}\bigg)- S(A|B)
 \notag\\
 &=&
 S \bigg( \tr_B \frac{(I\ox\p_0) \r
(I\ox\p_0)}{p_0}\bigg)- S(A|B)
 \notag\\
 &=&
 S \bigg(\sum^1_{k=0} p_k  \tr_B \frac{(I\ox\p_k) \r
(I\ox\p_k)}{p_k}\bigg)- S(A|B)
 \notag\\
 &=&
 S(\r_A)-S(A|B)
 \notag\\
 &=& I(\r).
 \eea
The second equality follows from the formula $p_0+p_1=1$, and the
fourth equality from Eq. \eqref{ea:p0+p1}. It follows from Eq.
\eqref{ea:discord} that Eq. \eqref{ea:necessity} holds only if
$C(\r)=0$. Then Lemma \ref{le:C(rho)=0} implies that $\r$ is a
product state, so $(e)\ra(d)$ follows.

\textit{(d)$\ra$(f)}. The hypothesis $I(\r)=0$ implies that $\r$ is
a product state. So the discord is also zero and the assertion
follows.

\textit{(f)$\ra$(g)}. It is a corollary of Eq.
\eqref{ea:super>=normal}.

\textit{(g)$\ra$(a)}. By Eq. \eqref{ea:superDISCORD} and the
concavity of von Neumann entropy, we have
 \bea
 \label{ea:(g)ra(a)}
 D_w(\r)
 &=& \min_{ \{\p_0,\p_1\} } S_w (A|\{ P^B(x) \}) - S(A|B)
 \notag\\
 &=& S(\r_A) - S(A|B)
 \notag\\
 &\ge& \max_{ \{\p_0,\p_1\} } S_w (A|\{ P^B(x) \}) - S(A|B).
 \eea
So the quantity $S_w (A|\{ P^B(x) \})$ is constant for any
$\p_0,\p_1$. The equality in Eq. \eqref{ea:(g)ra(a)} holds if and
only if $\r_{ A | P^B(x) }=\r_{ A | P^B(-x) }=\r_A$ for any
$\p_0,\p_1$. By Eqs. \eqref{ea:APB(pm x)}, \eqref{ea:P(x)} and
\eqref{ea:p0+p1}, we have $(I\ox\pi_k) \r (I\ox\pi_k)\propto\r_A$.
This fact and Eq. \eqref{ea:discord} imply that $D(\r)=I(\r)$. So we
have proved (g)$\ra$(f)$\ra$(b)$\ra$(a), where the last relation
follows from Lemma \ref{le:C(rho)=0}. This completes the proof.
 \epf

As a typical example, Theorem \ref{thm:zeroSUPERdiscord} implies
that the completely mixed state $\frac14 I\ox I$ has zero super
discord. This observation has been included as a special case of
\cite{sp12}. Note that the equivalence in Theorem
\ref{thm:zeroSUPERdiscord} does not hold for states with zero
discord, because such states may be not product states.

Super discord with weak measurement has a nature analogue with
Gaussian quantum discord [37] restricted to Gaussian measurements in
the realm of continuous-variable (CV) systems, where the measurement
class is larger than the class running over all local projective
measurements. Recently, quantum discord with non-Gaussian
measurements has been also studied [38]. Theorem 1 tells us that
weak measurement can reveal much correlation in the sense that super
discord is always larger than quantum discord and vanishing only on
product state. Therefore super discord is ubiquitous in quantum
system. A natural and interesting question is that can Gaussian
measurements or even non-Gaussian measurements also reveal much more
quantum correlation than the sole projections? We propose it as an
open question, as it may help us obtain a deep understanding of how
the quantum correlation behaviors with regards to different
measurement classes.


 It is widely accepted that mutual information contains both
classical correlation and quantum correlation. However, they share
the same vanishing condition with super discord. By Theorem
\ref{thm:zeroSUPERdiscord}, we can also say that super discord is
larger than normal discord generally, which makes that super discord
really likely reveal much more quantum correlation than quantum
discord. Another superiority of super discord is that their
vanishing does \emph{not} rely on the specific side, although we
have the measurement acting on ``left'' or ``right'' system. The
reason is if ``left'' super discord is zero, then the state is a
product state by Theorem \ref{thm:zeroSUPERdiscord}. So the
``right'' super discord must be also zero. As an application, we
illustrate how super discord present in optimal assisted
discrimination that different from normal discord in next section.

\section{ super discord in optimal state discrimination}\label{sudiscrimina}

Super discord vanishes only on product states. It exists more widely
in quantum information processing than other quantum correlation
such as entanglement and quantum discord. That is, one can find some
assignments in which super discord is nonzero, while neither of
entanglement and quantum discord is nonzero. In this section, we
consider state discrimination scheme and provide an example in which
super discord exists more widely than entanglement and quantum
discord. In order to achieve this goal, we will focus on the case of
the minimal error probability, where entanglement and one side
quantum discord vanish in the scheme.


We first review the scheme of state discrimination introduced by
Roa, Retamal and Alid-Vaccarezza (RRA scheme) \cite{Roa}.  Consider
two nonorthogonal states $|\psi_+\rangle$ and $|\psi_-\rangle$
 is randomly prepared in one of the  \emph{priori} probabilities
$p_+$ and $p_-$ with $p_++p_-=1$. To discriminate the two states $|\psi_+\rangle$ or $|\psi_-\rangle$.
 Couple the original system to an auxiliary qubit $A$ by a joint unitary
transformation $U$ such that
\begin{eqnarray}
U|\psi_+\rangle|k\rangle_a=\sqrt{1-|\alpha_+|^2}|+\rangle|0\rangle_a+\alpha_+|0\rangle|1\rangle_a, \nonumber\\
U|\psi_-\rangle|k\rangle_a=\sqrt{1-|\alpha_-|^2}|-\rangle|0\rangle_a+\alpha_-|0\rangle|1\rangle_a,
\label{unitary}
\end{eqnarray}
where $|k\rangle_a$ is an auxiliary state with orthonormal basis $\{|0\rangle_a,|1\rangle_a\}$,
$|\pm\rangle\equiv (|0\rangle\pm|1\rangle)/\sqrt{2}$ are the orthonormal states of the system
that can be discriminated.

The state of the system and ancilla qubits is now given as
\begin{eqnarray}\label{rho}
\rho_{|\alpha_+|} & = & p_+U\left( |\psi_+\rangle\langle\psi_+|\otimes |k\rangle_a \langle k|\right) U^\dag\nonumber\\
&  & + p_-U\left( |\psi_-\rangle \langle\psi_-|\otimes |k\rangle_a \langle k|\right) U^\dag.
\end{eqnarray}
It is shown that the conclusive recognition between two
nonorthogonal states relies on the existence of entanglement and
discord in general case of RRA scheme \cite{Roa}. However, in the
optimal case or with maximum recognized probability, only one side
(``right'' side) discord (or dissonance) is nonzero \cite{bo1}. An
interesting question is that what kinds of non-classical correlation
can be regarded as a candidate for resource for the scheme in the
optimal case? By the fact that the super quantum discord is always
greater than or equal to the normal quantum discord and the
equivalent condition given in Theorem \ref{thm:zeroSUPERdiscord}, we
guess that the super quantum discord really catch the non-classical
correlation of the RRA scheme on both side.

\begin{figure}\centerline{
\includegraphics[width=8cm]{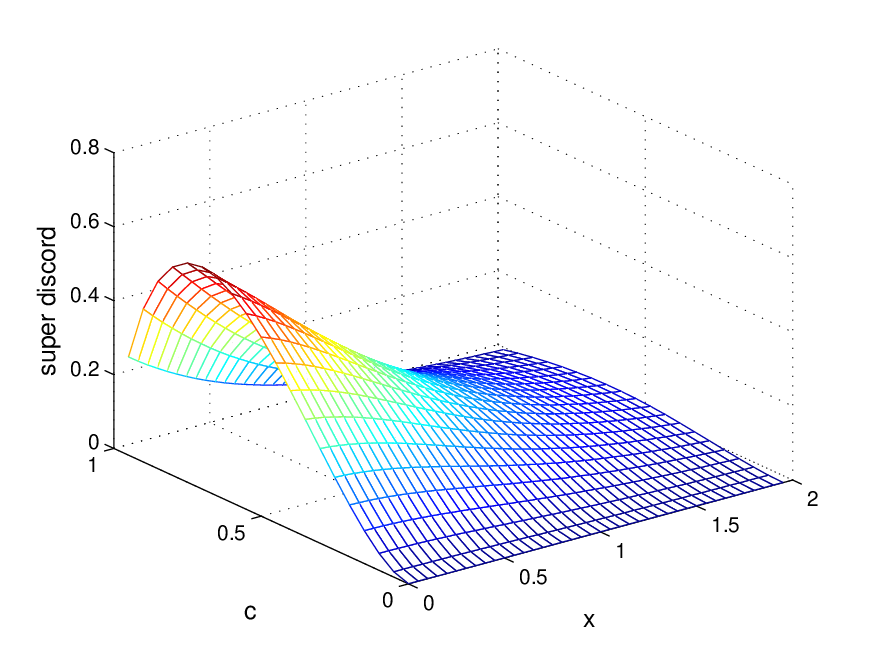}}
\caption{ super discord in the optimal case of assisted state discrimination
as a function of $\alpha_+=c$, and the  strength $x$ in the measurement process
for $0\leq\alpha_+\leq 1$ and $0\leq x\leq 2$.
}\label{ass}
\end{figure}

In the following, we concentrate on the state (\ref{rho}) in zero
``left'' discord cases. Since ``right'' discord is always presented
in the quantum system, and super discord is larger than normal
discord, we have concluded that the ``right'' super discord must be
nonzero in the scheme. As is shown in \cite{bo1}, in the cases when
the ``left'' discord disappears, the following three items must be
satisfied: $\alpha$ is a real number, and $\alpha\geq 0$;
$p_+=p_-=\frac{1}{2}$;
$|\alpha_+|=|\alpha_-|=\sqrt{|\alpha|}=\sqrt{\alpha}$ and the case
is indeed  the optimal assisted state discrimination. For
convenience, we set $\alpha_+=c$ be a real number. Then the state
$\rho$ in (\ref{rho}) is reduced to
 \begin{eqnarray}\label{rhoopt}
\rho_{c} & = & \frac{1-c^2}{2}(I\otimes |0\rangle \langle 0|)+|0\rangle \langle 0|\otimes\nonumber\\
&  & [c^2|1\rangle \langle 1|+\frac{\sqrt{2}c\sqrt{1-c^2}}{2}(|0\rangle \langle 1|+|1\rangle \langle 0|)].
\end{eqnarray}
We use the weak measurement $P(x)\otimes I, P(-x)\otimes I$ to act
the state $\rho_{c}$ in (\ref{rhoopt}), where $P(x), P(-x)$ is given
by Eqs. \eqref{ea:P(x)} and \eqref{ea:P(-x)}, and
$\pi_0=|\psi\rangle\langle\psi|,\pi_1=|\widetilde{\psi}\rangle\langle\widetilde{\psi}|$,
$|\psi\rangle=\cos\theta|0\rangle+e^{i\varphi}\sin\theta|1\rangle$,
$|\widetilde{\psi}\rangle=\sin\theta|0\rangle-e^{i\varphi}\cos\theta|1\rangle$.
Then the weak ``left'' conditional entropy for this state is given
by
\begin{eqnarray}\label{conditionentropy}
& &S_w(B|\{P^A(x)\})  =  -p(x)[\lambda_+(x)\log\lambda_+(x)\nonumber\\
&  & +\lambda_-(x)\log\lambda_-(x)]-p(-x)[\lambda_+(-x)\log\lambda_+(-x)\nonumber\\
& &+\lambda_-(-x)\log\lambda_-(-x)],
\end{eqnarray}
here $p(x)=\frac{1}{2}(1-\tanh(x)\cos(2\theta)c^2)$, and,
\begin{eqnarray}
\lambda_\pm(x)&=&\frac{1}{2(1-\tanh(x)\cos(2\theta)c^2)}(1-\tanh(x)\cos(2\theta)c^2
\nonumber \\
&&\pm(1-2c^2+2c^4-2c^2\tanh(x)\cos(2\theta)
\nonumber \\
&&+(2c^2-c^4)(\tanh(x)\cos(2\theta))^2)^{\frac {1}{2}}),
\end{eqnarray}
and $\lambda_\pm(-x)$ can be similarly defined. After calculation, we find that $S(AB)=S(A)$.
Let $D_w(B:A)=D_w(\r)$ in Eq. \eqref{ea:superDISCORD} by exchanging
systems A and B. From Eqs. \eqref{ea:superDISCORD} and
\eqref{conditionentropy} we have
\begin{eqnarray}\label{SUR}
 D_w(B:A)=\min_{\{\p_i^A\}}S_w(B|\{P^A(x)\})=\min_{\theta}S_w(B|\{P^A(x)\}),
\end{eqnarray}
and $D_w(B:A)$ is a function of $x$ and $c$. In Fig. \ref{ass}, we
have plotted the picture of $D_w(B:A)$. We can see that for all
$0<c<1$, the super discord increases with the decreasing of the
strength of the measurement $x$. When $x\rightarrow +\infty$, the
weak measurement reduces to the strong measurement and the super
discord approach to normal discord. The discord and entanglement are
always zero in this optimal case. Thus, we have shown that super
discord can be regarded as a resource in optimal assisted state
discrimination.

In summary, we have found that in the optimal assisted
discrimination scheme, entanglement and one-sided quantum discord
are totally unnecessary. In this scheme we only need the super
discord, which always exists between the principal qubit and the
ancilla. It reveals the mysterious properties of non-classical
correlation in quantum information processing, and neither quantum
discord nor entanglement is the essential ingredient in
non-classical correlation. Our findings could stimulate more
research on the role of nonclassical correlation in quantum
information processing.

\section{Summary and discussions}\label{summ}

In this paper, we have obtained several equivalent conditions for
zero super discord. It is shown that the vanishing of super quantum
discord is equivalent to vanishing of classical correlation, mutual
information \emph{etc.} So super quantum discord is a kind of
quantum correlation which ubiquitously exists in quantum system.
Further more, super quantum discord can be present in some quantum
information processing tasks where entanglement is totally not
necessary and only one side quantum discord is nonzero.

One fundamental problem in quantum information is to quantify correlations.
Quantum discord emerges in separating total correlations into quantum and
classical parts. Some evidences show that all correlations behave as if
they were exclusively quantum \cite{Bennett2011,Ferraro,Pankowski}.
In this paper, we confirm this concept by showing that super quantum
discord vanishes only when mutual (total) information vanishes.
This extends the regime of quantumness of correlations to all bipartite quantum
states except the product state. With this conclusion, we may always refer
to quantum correlation in various quantum information protocols since
non-zero total correlation implies non-zero super quantum discord. With this
in mind, we may safely say that quantum correlation exists
in tiny signal amplifying processes where weak measurement is performed.

\acknowledgements

This work was carried out when L.C. was visiting the Institute
of Physics, CAS, China. He was mainly supported by MITACS and NSERC.
The CQT is funded by the Singapore MoE and the NRF as part of the
Research Centres of Excellence programme. B.L. and H.F. were supported by
``973'' program (2010CB922904),  NSFC (11175248,11305105) and NSFJXP (20132BAB212010).


\begin{thebibliography}{99}
\bibitem{neumann} J. von Neumann, \emph{Mathematische Grundlagen der
Quantenmechanik} (Springer, Berlin, 1932)
\bibitem{Nielsen&Chuang} M. A. Nielsen and I.~L. Chuang, \emph{Quantum Computation
and Quantum Information} (Cambridge University Press,
Cambridge, England, 2000).
\bibitem{aav} Y. Aharonov, D. Z. Albert, and L. Vaidman, Phys. Rev. Lett. {\bf 60}, 1351 (1988).
\bibitem{Korotkiv2006} A. N. Korotkov and A. N. Jordan, \prl {\bf97}, 166805 (2006).
\bibitem{Sun2009} Q. Sun, M. Al-Amri, and M. S. Zubairy, \pra {\bf80}, 033838 (2009).
\bibitem{Ueda1992} M. Ueda and M. Kitagawa, \prl {\bf68}, 3424
(1992);M. Koashi and M. Ueda, \prl {\bf82}, 2598 (1999).


\bibitem{brun} O. Oreshkov and T. A. Brun, Phys. Rev. Lett. {\bf 95}, 110409 (2005).

\bibitem{wise1} H. M. Wiseman, Phys. Lett. A {\bf 311}, 285 (2003).

\bibitem{mir}  R. Mir, J.S. Lundeen, M.W. Mitchell, A.M. Steinberg, J.L. Garretson, and
H.M. Wiseman, New J. Phys. {\bf 9}, 287 (2007).

\bibitem{Korotkov2010} A. N. Korotkov and K. Keane, \pra {\bf81}, 040103(R)
(2010);  Y.-S. Kim, J.-C. Lee, O. Kwon and Y.-H. Kim, Nature Physics {\bf8},
117 (2011). Y.-W. Cheong and S.-W.  Lee, \prl {\bf109}, 150402 (2012).

\bibitem{kwiat}O. Hosten and P. Kwiat, Science {\bf 319}, 787 (2008).

\bibitem{lundeen}J. S. Lundeen,B. Sutherland, A. Patel, C. Stewart and  C. Bamber\ Nature {\bf 474},
188 (2011).

\bibitem{hv01} L. Henderson and V. Vedral,  \jpa {\bf34}, 6899 (2001).


\bibitem{oz02} H. Ollivier and W. H. Zurek, \prl {\bf88}, 017901
(2001).

\bibitem{modi} K. Modi, A. Brodutch, H. Cable, T. Paterek, and V. Vedral,  Rev. Mod. Phys. {\bf84}, 1655 (2012)
\bibitem{Knill} E. Knill and R. Laflamme, Phys. Rev. Lett. {\bf 81}, 5672
(1998);  A. Datta, A. Shaji, and C. M. Caves, Phys. Rev. Lett. \textbf{100}, 050502 (2008).
\bibitem{dak} B. Daki{\'c},
Y. O. Lipp, X. Ma, M. Ringbauer, S. Kropatschek, S. Barz, T. Paterek, V. Vedral, A. Zeilinger, \v{C}. Brukner and P. Walther,
 Nature Physics {\bf 8}, 666 (2012).

\bibitem{gu} M. Gu, H. M. Chrzanowski, S. M. Assad, T. Symul, K. Modi, T. C. Ralph, V. Vedral, P. K. Lam, Nature Physics  {\bf 8}, 671 (2012).



\bibitem{Roa} L. Roa, J. C. Retamal, and M. Alid-Vaccarezza, Phys. Rev. Lett. {\bf 107}, 080401 (2011).
\bibitem{bo1} B. Li,  S Fei, Z Wang and H Fan, Phys. Rev. A \textbf{85}, 022328 (2012).
\bibitem{fulin} F. L. Zhang, J. L. Chen, L. C. Kwek, V. Vedral, Sci. Rep. {\bf3}, 2134(2013).
\bibitem{hhh96} M. Horodecki, P. Horodecki, and R. Horodecki,
Phys. Lett. A {\bf223}, 1 (1996).

\bibitem{peres1} A. Peres, Phys. Rev. Lett.  {\bf 77}, 1413 (1996);  K. Chen, S. Albeverio, and S. M. Fei, Phys. Rev. Lett. {\bf 95}, 040504 (2005)

\bibitem{cd12} L. Chen and D.\v{Z}. Djokovi{\'c}, Commun. Math. Phys. {\bf323}, 241(2013).

\bibitem{datta08} A. Datta, quant-ph/0807.4490 (2008).

\bibitem{dvb10} B. Daki{\'c}, V. Vedral, and \v{C}. Brukner, \prl {\bf105},
190502 (2010).

\bibitem{ccm11} L. Chen, E. Chitambar, K. Modi, G. Vacanti, \pra {\bf83}, 020101(R) (2011).

\bibitem{Bennett2011}C. H. Bennett, A. Grudka, M. Horodecki, P. Horodecki, and R. Horodecki,
Phys. Rev. A {\bf 83}, 012312 (2011).




\bibitem{sp12} U. Singh and A. Pati, quant-ph/1211.0939 (2012).

\bibitem{dd} D. Deutsch, Phys. Rev. Lett. {\bf 50}, 631 (1983).
\bibitem{ren} M. Berta, M. Christandl, R. Colbeck, J. M. Renes, and R. Renner, Nature Phys. {\bf 6}, 659 (2010).





\bibitem{ll11} N. Li and S. Luo,  \pra {\bf84}, 042124
(2011).

\bibitem{luo08} S. Luo, \pra {\bf77}, 042303
(2008).

\bibitem{lwf11} B. Li, Z. X. Wang, and S. M. Fei, \pra {\bf83}, 022321 (2011).
\bibitem{Ferraro} A. Ferraro and M. G. A. Paris, Phys. Rev. Lett. {\bf 108},  260403  (2012).

\bibitem{Pankowski}
{\L}ukasz Pankowski and Barbara Synak-Radtke, J. Phys. A: Math. Gen. \textbf{41}, 570308
 (2008).

 \bibitem{paolo} P. Giorda, and M. G. A. Paris, \prl {\bf105}, 020503 (2010).

 \bibitem{paolo1} P. Giorda, M. Allegra, and M. G. A. Paris, \pra {\bf86}, 052328 (2012).




\end{thebibliography}
\end{document}